\documentclass[sigconf,nonacm,screen]{acmart}

\RequirePackage{microtype}
\DisableLigatures{encoding = T1, family = tt* }

\usepackage[T1]{fontenc}
\usepackage[utf8]{inputenc}
\usepackage{url}
\usepackage[autolanguage]{numprint}
\usepackage{booktabs}
\usepackage{graphicx}
\usepackage{wasysym}
\usepackage{fancyvrb}
\usepackage{fvextra}
\usepackage{setspace}
\usepackage{tikz}
\usepackage{cprotect}
\usepackage{float}
\usepackage{adjustbox}
\usepackage{enumitem}

\usepackage{xspace}

\newcommand{\num}[1]{\numprint{#1}}
\newcommand{\per}[1]{\numprint[\%]{#1}}

\newenvironment{CVerbatim}
  {\center\BVerbatim}
  {\endBVerbatim\endcenter}

\newcommand{\alias}[2]{{\texttt{#1} $\rightarrow$ \texttt{#2}}}
\newcommand{\cmd}[1]{{\texttt{#1}}}

\begin{document}

\title{An Empirical Investigation of Command-Line Customization}

\author{Michael Schröder}
\orcid{0000-0003-1496-0531}
\affiliation{%
	\institution{TU Wien}
	\city{Vienna}
	\country{Austria}
}
\email{michael.schroeder@tuwien.ac.at}

\author{Jürgen Cito}
\affiliation{%
	\institution{TU Wien}
	\city{Vienna}
	\country{Austria}
}
\affiliation{%
	\institution{Massachusetts Institute of Technology}
	\city{Cambridge}
	\country{U.S.A.}
}
\email{juergen.cito@tuwien.ac.at}

\begin{abstract}
	The interactive command line, also known as the shell, is a prominent mechanism used extensively by a wide range of software professionals (engineers, system administrators, data scientists, etc.). Shell customizations can therefore provide insight into the tasks they repeatedly perform, how well the standard environment supports those tasks, and ways in which the environment could be productively extended or modified.
	To characterize the patterns and complexities of command-line customization, we mined the collective knowledge of command-line users by analyzing more than 2.2~million shell alias definitions found on GitHub.
	Shell aliases allow command-line users to customize their environment by defining arbitrarily complex command substitutions.
	Using inductive coding methods, we found three types of aliases that each enable a number of customization practices: 
	\textsc{Shortcuts} (for \emph{nicknaming commands}, \emph{abbreviating subcommands}, and \emph{bookmarking locations}),
	\textsc{Modifications} (for \emph{substituting commands}, \emph{overriding defaults}, \emph{colorizing output}, and \emph{elevating privilege}),
	and \textsc{Scripts} (for \emph{transforming data} and \emph{chaining subcommands}).
	We conjecture that identifying common customization practices can point to particular usability issues within command-line programs, and that a deeper understanding of these practices can support researchers and tool developers in designing better user experiences.
	In addition to our analysis, we provide an extensive reproducibility package in the form of a curated dataset together with well-documented computational notebooks enabling further knowledge discovery and a basis for learning approaches to improve command-line workflows.
\end{abstract}

\keywords{command line, customization practices, collective knowledge, inductive coding}

\settopmatter{printfolios=true}
\maketitle

\section{Introduction}

A command-line interface, also called a \emph{shell}, is a textual interface that allows users to interact with the underlying operating system by issuing commands.
Expert users, such as system administrators, software developers, researchers, and data scientists, routinely use the shell as it affords them flexibility and the ability to compose multiple commands.
They perform a variety of tasks on their systems including navigating and interacting with the filesystem (e.g., \cmd{ls}, \cmd{mv}, \cmd{cd}), using version control (e.g., \cmd{git}, \cmd{hg}), installing packages (e.g., \cmd{apt-get}, \cmd{npm}), or dealing with infrastructure (e.g., \cmd{docker}).
Experts can adapt and play with a multitude of commands and arguments, chaining them together to create more complex workflows.
All this versatility introduces a common problem in user interfaces of recognition over recall~\citep{nielsen:05}, where users have to recall the particularities of syntax and argument combinations, instead of enabling them to use a more recognizable symbol (as in graphical user interfaces).

A way for these experts to introduce recognizability and customize their command-line experience is to attach distinct names to potentially convoluted, but frequently used, command and argument structures, as well as workflows expressed as compositions of commands.
This can be achieved by defining shell aliases.
An alias substitutes a given name, the \emph{alias}, with a string value that defines an arbitrarily complex command (or chain of commands).
The set of aliases users define provides a window into their preferences expressed as part of their personal configuration.
Many users publicly share these configurations on social coding platforms such as GitHub, contributing to a collective knowledge of command-line customizations, which can provide insight into the tasks that expert users repeatedly perform and how well the standard environment supports those tasks. 

\subsection{Contribution}

We see our large-scale analysis on command-line user customizations manifested in alias definitions as a unique window of opportunity to study how the standard environment of the command line could be productively extended, modified, and improved. 
Our work goes hand in hand with existing efforts to innovate on the experience of command lines that employ techniques from research in systems~\citep{posh, odfg}, software engineering and programming languages~\citep{pash, kumquat, dantoni:17}, human-computer interaction~\citep{bespoke:19, gandhi2020}, and artificial intelligence~\citep{clai, nl2bash, findcmd}.
Particularly, our extensive qualitative and quantitative analysis, in conjunction with our dataset, form the basis for identifying opportunities for improving command-line experience in the following directions: by characterizing customization practices, we gain a categorical understanding underlying the needs and wants of command-line users; based on our analysis, we identify opportunities for innovation and formulate them as implications, accompanied with concrete scenarios and examples; % that aim to support command-line users, tool developers, and the design of the Shell in general. 
further, our comprehensive dataset enables the foundation of learning approaches, as part of learning-based program synthesis \citep{bruch:09,raychev2014completion}, automated repair \citep{monperrus:18}, and recommendation systems~\citep{mens2014source};
finally, we also see our results and datasets as a basis for usability research that can impact the design of tools and the future of the shell in general.

%Consequently, we performed an exploratory analysis on the population of alias definitions on GitHub. 
%Improving the shell holds much promise for development, ops, and data processing
%To study these preferences, we performed an exploratory analysis on the population of alias definitions on GitHub. 

We summarize the work in this paper as follows:

\begin{itemize}

	\item We identified nine \textbf{Customization Practices}, grouped into three high-level themes:
	\textsc{Shortcuts} introduce new names.
	They can be used for \emph{nicknaming commands} (and correcting misspellings in the process),
	\emph{abbreviating subcommands} like \texttt{git push},
	and \emph{bookmarking locations} for quick navigation.
	\textsc{Modifications} change the semantics of commands.
	We can use these types of aliases for \emph{substituting commands}, such as replacing \cmd{more} with \cmd{less},
	for \emph{overriding defaults} to customize commands to personal contexts, 
	which often involves \emph{colorizing output},
	and also running certain commands as root by \emph{elevating privilege}.
	Aliases that combine multiple commands are \textsc{Scripts}.
	They enable many ways of \emph{transforming data} using Unix pipes, 
	and allow for automating repetitive workflows by \emph{chaining subcommands}.
	\smallskip
	
	\item A \textbf{Curated Dataset of Command-Line Customizations}, consisting of over 2.2 million shell aliases collected from GitHub.
	We view our dataset as a playground for fine-grained discovery that can benefit researchers, tool-builders, and command-line users;
	for example, researchers can use this knowledge base to discover which commands are frequently used together and how they are combined, while tool-builders can see how their programs are being customized.
We also describe the effective mining technique we used to distill this knowledge, which allowed us to capture almost the whole population (\per{94.09}) of relevant shell configuration files.
	\smallskip	
	
	\item We formulate \textbf{Implications for Improving Command-Line Experience} that go beyond single customization practices to address shortcomings and tie them to existing user experience research.
	Codifying emergent behavior~\citep{fast:14} found in our customizations enables \emph{learning repair rules} and \emph{discovering workflows}.
	We are able to \emph{uncover conceptual design flaws}, where customizations indicate frustrations with underlying command structures, supporting prior research on potential flaws in the conceptual design of certain commands~\citep{perez:13}.
	Based on the prevalence of highly variable command redefinitions, we propose \emph{contextual defaults}, the ability to suggest different command preferences based on user context~\citep{stefanidis:11}.
	Overall, we find that many customizations deal with the tension of \emph{Interactivity vs Scripting}: commands being used to interactively navigate systems, while at the same time being used within scripts for batch-processing.
\end{itemize}
We now describe usage and syntax of aliases as a vehicle for customization. We further describe our data collection and coding process, followed by a presentation of customization practices. Finally, we discuss implications for usability and review related work in the broader context of this study.

\section{Background}

A shell is a command interpreter allowing the user to interact with an underlying system.
The concept of the operating system shell as an independent process executing outside the kernel originated in Multics \citep{pouzin:65} and was further developed into the original Unix shell \texttt{sh} and its various descendants \citep{jones:11,seibold2020shell}.
The POSIX family of standards defines a Shell Command Language~\citep{posix_standard, greenberg:17}, whose standard implementation is still the \texttt{sh} utility, but there exist a wide variety of popular POSIX-compliant shells like \texttt{bash} or \texttt{zsh}.
These implementations are free to extend the functionality of the shell, but all share a common subset of core commands and programming language constructs.
In this paper, we focus on the built-in \texttt{alias} command, available on all POSIX shells.

\subsection{Usage and Syntax}

The \texttt{alias} command allows the user to create \emph{alias definitions}, defining command substitutions.
When the shell processes the command line, it replaces known alias names with their defined string values.
For example, 
\begin{CVerbatim}
alias ll='ls -l'
\end{CVerbatim}
defines the \emph{alias name} \texttt{ll}, that is replaced by the \emph{alias value} \texttt{ls -l}.
In this case, \cmd{ls} is the standard command for listing directory contents, with the argument \texttt{-l} specifying a long-form output format.
So the alias \texttt{ll} (present in many system configurations) is used to specify a default argument to a commonly used command under a different name.

Alias values can be arbitrarily complex strings and can substitute not only simple commands and arguments, but whole chains of commands. 
The definition
\begin{CVerbatim}
alias ducks='du -cksh * | sort -hr | head -n 15'
\end{CVerbatim}
defines the new command \cmd{ducks} by chaining together three different command-line tools in order to return the 15 largest files in the current directory.

%alias ip=ifconfig | grep "inet " | grep -v 127.0.0.1 | cut -d' ' -f2
%which chains together several commands to find out the IP addresses of the system.

%alias unwip="git log -n 1 | grep -q -c wip && git reset HEAD~1"
%which chains together three commands to rewind the most recent commit in a git repository, if its commit message contains the word "wip".

In general, an alias definition takes the form
\begin{CVerbatim}
alias name=value
\end{CVerbatim}
where \verb|value| can optionally be enclosed in single (\verb|'|) or double (\verb|"|) quotes and \verb|name| can be any identifier that is a valid command name.\footnote{Some shells allow for an alternative alias syntax without the equals sign between \texttt{name} and \texttt{value}. In this paper we only look at POSIX-compliant alias definitions.}
% TODO: technically not quite correct, see FSE review 2
In particular, the alias name can be an existing command, so a re-definition like
\begin{CVerbatim}
alias grep='grep --color=always'
\end{CVerbatim}
is possible.

In the remainder of this paper, we will use the more compact notation \alias{a}{b} to indicate an alias that replaces the name \texttt{a} with the value \texttt{b}.

\subsection{Dotfiles}

Aliases can be entered directly on the command line, in which case they are valid until the shell session ends.
To make an alias definition permanent, it is common practice to enter it into a file that is read and executed by the shell on startup.
The names of these configuration files differ by shell, but common ones are \verb|.bashrc|, \verb|.zshrc|, or \verb|.profile| and their main difference is the order in which they are executed.\footnote{\url{https://www.gnu.org/software/bash/manual/html_node/Bash-Startup-Files.html} or \url{https://zsh.sourceforge.io/Doc/Release/Files.html}}
Often, aliases are also stored in other files referred to by these startup scripts.

These kinds of files---text-based configuration files that store system or application settings---are also known as \emph{dotfiles}, because their filenames usually start with a dot (\verb|.|) so that they are hidden by default on most Unix-based systems.
In recent years, people have started sharing their dotfiles on platforms like GitHub.\footnote{\url{https://dotfiles.github.io}}
This has the advantage of being able to sync one's configurations across different machines, and also enables exchange and discovery of configurations between users.

\section{Dataset}

Our analysis is based on \num{2204199} alias definitions found on GitHub, collected over a period of two-and-a-half weeks from December 20th 2019 to January 8th 2020.

\subsection{Data Collection}

Alias definitions can appear in any Shell script, but we anticipated that they would predominantly be found in personal configuration files (like \verb|.bashrc| or \verb|.bash_profile|).
%While we did not want to focus exclusively on these personal configuration scripts, as we want to study the wide range of different alias uses, it is important that they are appropriately accounted for in order for our data to be representative.
Unfortunately, this rules out using some prominent existing datasets for our study~\citep{mombach}:
The public GitHub archive on BigQuery,\footnote{\url{https://bigquery.cloud.google.com/table/bigquery-public-data:github_repos.files}} while containing over 1.5~TB of source code, only includes ``notable projects'' (presumably those with a certain number of stars on GitHub) that additionally have an explicit open source license. 
This leaves out many of the repositories we are interested in, as users sharing configuration scripts for personal use do not usually add a license file and their repositories are generally not ``notable''.
GHTorrent~\citep{ghtorrent}, another popular archive of GitHub data, only contains metadata but not file contents.

Therefore, we found it necessary to write our own tooling to directly collect the data from GitHub ourselves.
We used the GitHub Code Search API\footnote{\url{https://docs.github.com/en/rest/reference/search}} to find files written in Shell language\footnote{GitHub uses the Linguist library to classify code: \url{https://github.com/github/linguist}} that contain the string \verb|alias|.

Alas, the GitHub Code Search API comes with its own set of limitations:
\begin{enumerate}
    \item only files smaller than 384 KB are searchable
    \item forks are not included
    \item requests are rate limited at 30 per minute and there are additional opaque abuse detection mechanisms that impose further restrictions in an unforeseeable manner
    \item the number of results is limited to \num{1000} per search request
\end{enumerate}
The first two limitations do not really affect us, as we are interested in smaller files and do not have to consider forks.
The rate limiting, while significantly slowing down the retrieval process, is also not a fatal obstacle.
The maximum number of returned search results, however, is a critical limitation.
To get around it, we wrote a Python tool called \verb|github-searcher|\footnote{\url{https://github.com/ipa-lab/github-searcher}} that uses a clever sampling strategy to vastly increase the number of results we are able to retrieve.

The sampling strategy is based on the GitHub API allowing code search queries to be conditioned on file sizes. 
For example, the query 
\begin{CVerbatim}
alias language:Shell size:101..200
\end{CVerbatim}
returns up to \num{1000} Shell language files containing the string ``alias'' that have a file size between 101 and 200 bytes (inclusive).
Repeating the search with 
\begin{CVerbatim}
alias language:Shell size:201..300
\end{CVerbatim}
returns up to \num{1000} files of a size between 201 and 300 bytes, and so on.
Repeatedly searching with the same search term but different non-overlapping file size ranges allows us to significantly increase our sample of the overall population.
Another trick further improves on this: 
the API gives us an option to sort the results by most or least recently indexed;
if we run a search using a specific sort order, then we can effectively double the sample size by repeating the same search with the opposite sort order.
Thus we can get up to \num{2000} results per search per file size range.

Additionally, while GitHub does not allow us to retrieve more than a limited number of files per query, it does return the total count of files matching the query.
While this count is usually very erratic on broad searches, fluctuating wildly between repeated requests, it turns out to be fairly accurate for searches with a small number of results, such as those conditioned on a narrow range of file sizes.
This allows us to get a good estimate of the population, and how accurately our sample approximates it.

For this study, using the search term
\begin{CVerbatim}
alias language:Shell
\end{CVerbatim}
and the sampling strategy described above, we started by sampling all files in increments of 100 bytes and stopped when we reached 29~KB, about ten times the median file size of the estimated population encountered so far.
We then re-sampled some high-population areas with smaller size increments in order to get a better sample, in some cases sampling in increments of 1 byte.
In total, we collected \num{844140} files from \num{304361} GitHub repositories.
Our sample represents \per{94.09} of the estimated population of \num{897182} files under 29~KB on GitHub written in Shell language and containing the word ``alias''.
The file contents, together with repository metadata, were stored in an SQLite database.
After removing duplicate files based on their SHA-1 hash value, our database contains \num{372816} unique files from \num{205126} repositories.

\begin{figure}
    \centering
    \includegraphics[width=0.85\columnwidth]{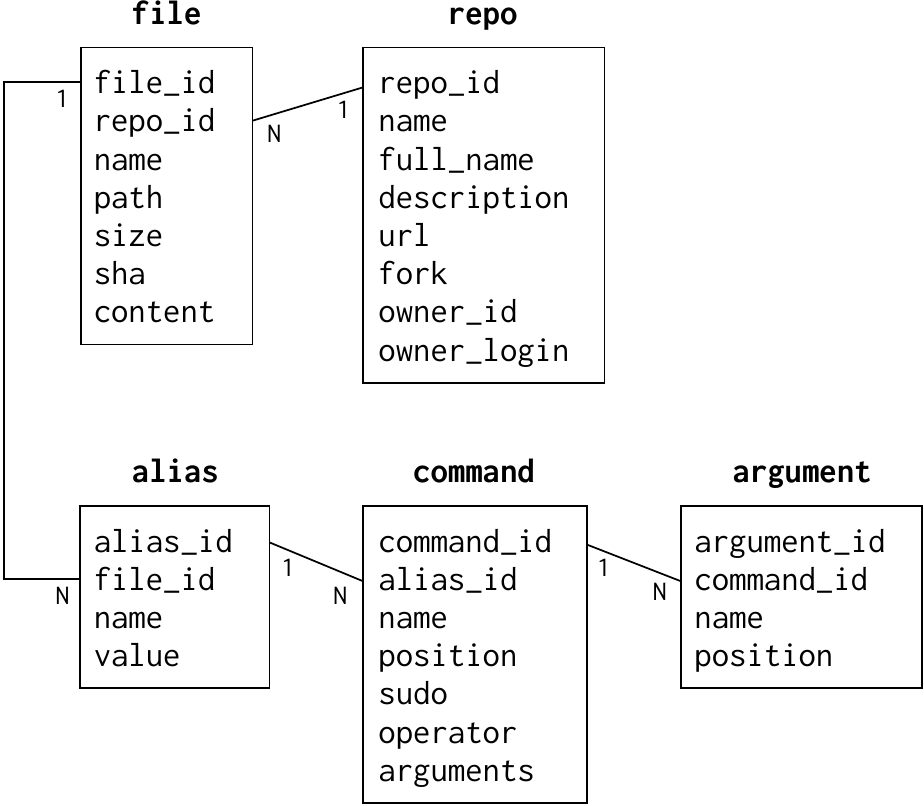}
    \caption{Relational database schema.}
    \label{fig:schema}
\end{figure}

\subsection{Parsing}

\begin{figure*}
	\centering
	\includegraphics[width=0.85\textwidth]{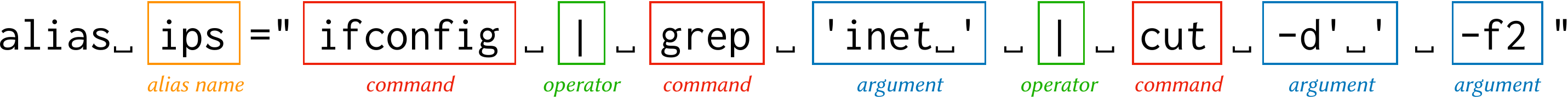}
	\cprotect\caption{Decomposition of \verb`alias ips="ifconfig | grep 'inet ' | cut -d' ' -f2"`.}
	\label{fig:parser}
\end{figure*}

After collecting files with potential aliases, we ran a parsing script to find actual alias definitions and decompose them into their constituent parts for analysis.
The decomposed aliases are stored in the same SQLite database as the raw file contents to facilitate easy cross-referencing.
The database schema is given in Fig.~\ref{fig:schema}.

The parser is a Haskell script that splits each alias definition into alias name and alias value, and tokenizes the value into commands and arguments.
Commands can be delimited by the shell operators for piping (\verb!|! and \verb!|&!), logical composition (\verb|&&| and \verb!||!), background execution (\verb|&|) and simple chaining (\verb|;|).
Arguments are separated by whitespace, but care is taken to handle quoted arguments correctly. 
For example, \verb|echo "hello world"| is parsed as one command (\texttt{echo}) with one argument (\texttt{"hello world"}).
See Fig.~\ref{fig:parser} for a more elaborate example.

Beyond quoting, which is defined by the Shell Command Language and thus uniform across all commands, the parser can not make any further considerations as to how arguments are meant to be interpreted.
While there are some conventions around command-line argument handling, programs are generally free to do as they wish and there is a wide variety of argument styles in the wild:
single-dash short arguments combined with double-dash long-form arguments (e.g., \texttt{ls -l -a --color=always});
combined short arguments without a dash (e.g., \texttt{tar xvzf archive.tar});
dictionary-style arguments (e.g., \texttt{dd if=/dev/zero of=/dev/sda});
subcommands (e.g., \texttt{git commit -m "wip"});
and many more.
Since the parser can not know the intentions of any command, it simply treats each token as a separate argument.
There is one exception: if the command is \cmd{sudo}, then its first argument is taken as the real command. 
For example, \texttt{sudo apt-get install} is parsed as the command \cmd{apt-get} with argument \texttt{install} and the sudo flag set.

After parsing, we ended up with \num{2204199} alias definitions, broken down into \num{2534167} commands and \num{3630423} arguments.
Files that did not contain any aliases were removed from the database, as was repository metadata that only referenced files without aliases.
\num{194218} files from \num{138112} repositories, or \per{52.09} of the original sample without duplicates, contained aliases.

\subsection{Provenance}

The majority of aliases in our dataset (\per{85.74}) originate from common startup scripts, like \texttt{.bashrc}, \texttt{aliases.zsh} or \texttt{.profile} (see Table~\ref{tab:file-names}).
We found another \per{2.78} of aliases originating from scripts related to Git, with file names like \texttt{git.plugin.zsh} or \texttt{git.bash}.
The remaining aliases are more or less evenly distributed among a variety of file names, none of which contributes more than half a percent of aliases, in most cases significantly less.
The average number of aliases per file is $11 \pm 18$, the median is 6.

\begin{table}
	\caption{Distribution of common file names}
	\label{tab:file-names}
	\begin{tabular}{rrlrr}
    \toprule
             \% &         Files &           Name Pattern &        Aliases &           \% \\
    \midrule
    \num{14.35} & \num{27870}  &                        \verb|*alias*| & \num{612516}  & \num{27.79} \\
    \num{27.72} & \num{53844}  &                       \verb|*bashrc*| & \num{591396}  & \num{26.83} \\
    \num{22.15} & \num{43011}  &                        \verb|*zshrc*| & \num{487002}  & \num{22.09} \\
     \num{9.42} & \num{18298}  &                      \verb|*profile*| & \num{199009}  &  \num{9.03} \\
     \num{1.26} &  \num{2455}  &                           \verb|git*| &  \num{61248}  &  \num{2.78} \\
    \bottomrule
\end{tabular}

\end{table}

\begin{table}
	\caption{Most common words in repository descriptions}
	\label{tab:repo-words}
	\begin{tabular}{rrlrr}
\toprule
         \% &       Repos & Word in Description &      Aliases &          \% \\
\midrule
\num{21.91} & \num{30259} &              \texttt{my} & \num{582448} & \num{26.42} \\
\num{17.00} & \num{23483} &        \texttt{dotfiles} & \num{466006} & \num{21.14} \\
\num{12.75} & \num{17612} &           \texttt{files} & \num{316963} & \num{14.38} \\
 \num{6.85} &  \num{9466} &   \texttt{configuration} & \num{175333} &  \num{7.95} \\
 \num{5.33} &  \num{7364} &          \texttt{config} & \num{131430} &  \num{5.96} \\
 \num{4.54} &  \num{6269} &        \texttt{personal} & \num{113885} &  \num{5.17} \\
 \num{4.13} &  \num{5707} &           \texttt{linux} & \num{101747} &  \num{4.62} \\
 \num{3.17} &  \num{4385} &            \texttt{bash} &  \num{94739} &  \num{4.30} \\
 \num{3.88} &  \num{5353} &         \texttt{scripts} &  \num{91021} &  \num{4.13} \\
 \num{2.06} &  \num{2840} &             \texttt{zsh} &  \num{74034} &  \num{3.36} \\
\bottomrule
\end{tabular}	
\end{table}

Table~\ref{tab:repo-words} shows the most commonly occurring words in repository descriptions on GitHub (excluding stop words), together with the amount of aliases found in repositories whose descriptions contain at least one of these words.
Counting them all together, repositories mentioning any of the words listed in Table~\ref{tab:repo-words}, in either their description or their repository name, make up \per{74.48} of the repositories in our dataset, contributing \per{82.3} of all aliases.
It is notable that more than half of the repositories in our dataset (\per{51.08}) have a name that includes the string \texttt{dot}, as in \texttt{dotfiles}, \texttt{dot-files}, \texttt{dots}, \texttt{mydotfiles}, and so on.
Looking at these names and descriptions, we can see a clear bias towards personal configurations and settings management.
On average, each repository contributes $16 \pm 28$ aliases, the median is 8.

\subsection{Reproducibility}
\label{sec:reproducibility}

To enable reproducibility and follow-up studies, we have made all data and our entire tool-chain publicly available.
Our dataset (1.45~GB of parsed alias definitions, plus 4.3~GB unparsed file contents and metadata) is available on Zenodo.\footnote{\url{https://doi.org/10.5281/zenodo.4007049}}
The parsing script and the executable Jupyter notebooks, containing all SQL queries and additional Python code used during our analysis, are available on GitHub.\footnote{\url{https://github.com/ipa-lab/shell-alias-analysis}}

\section{Analysis}

Table~\ref{tab:top-summary} shows the most common alias names, commands, and arguments appearing in alias definitions.
The most common alias name we found is \texttt{ls}, appearing a total number of \num{83782} times, which is \per{3.8} of all alias definitions.
Note that this is \texttt{ls} as an \emph{alias name}, a redefinition of the \texttt{ls} \emph{command}, which appears \num{260156} times (\per{10.27}).
This is a bit less often than \texttt{git}, the most common command, which appears in \num{327786} aliases (\per{12.93}).
The most common argument, across all commands, is \texttt{--color=auto}, appearing \num{153931} times (\per{4.24})

\begin{table*}
    \centering
    \caption{Top alias names, commands and arguments.}
    \label{tab:top-summary}
    \begin{tabular}{lrr}
\toprule
   Name &           \# &          \% \\
\midrule
    \verb|ls| &  \num{83782} &  \num{3.80} \\
    \verb|ll| &  \num{62465} &  \num{2.83} \\
  \verb|grep| &  \num{44479} &  \num{2.02} \\
    \verb|la| &  \num{43760} &  \num{1.99} \\
     \verb|l| &  \num{39539} &  \num{1.79} \\
\bottomrule
\end{tabular}
\hspace{0.2cm}
\begin{tabular}{lrr}
    \toprule
           Command &            \# &           \% \\
    \midrule
        \verb|git| &  \num{327786} &  \num{12.93} \\
         \verb|ls| &  \num{260156} &  \num{10.27} \\
         \verb|cd| &  \num{166632} &   \num{6.58} \\
       \verb|grep| &   \num{89598} &   \num{3.54} \\
        \verb|vim| &   \num{46545} &   \num{1.84} \\
    \bottomrule
\end{tabular}
\hspace{0.2cm}
\begin{tabular}{lrr}
    \toprule
                Argument &            \# &          \% \\
    \midrule
     \verb|--color=auto| &  \num{153931} &  \num{4.24} \\
               \verb|-i| &   \num{70640} &  \num{1.95} \\
               \verb|-a| &   \num{42910} &  \num{1.18} \\
               \verb|-l| &   \num{39519} &  \num{1.09} \\
               \verb|-v| &   \num{35295} &  \num{0.97} \\
    \bottomrule
\end{tabular}

\end{table*}

Looking at each part of an alias definition in isolation can only get us so far, as arguments only gain meaning in conjunction with commands and alias names can be identical between users, referring to the same command/argument combination, or indeed can overlap, meaning the same alias name is used differently by different users.
Table~\ref{tab:command-summary} gives a more informative view for the top two commands, \texttt{git} and \texttt{ls}, showing us the top arguments given with each and the most common alias names by which the command/argument combinations are referred to.
Here we can already identify some of the typical alias use cases.
Looking at \texttt{ls}, we find that aliases are used
to redefine the command with a default argument (\alias{ls}{ls --color=auto});
to shorten a common invocation (\alias{ll}{ls~-alF});
and to correct a spelling mistake (\alias{sl}{ls}).
We also notice that in the case of \texttt{git}, most aliases are used for shortening \texttt{git} subcommand invocations (e.g. \alias{gd}{git diff}).

\begin{table}
    \centering
    \caption{Top two commands with top arguments and aliases. %How to read this table: The most common argument for \cmd{git} is \texttt{status}, occurring in \per{5.85} of all aliases mentioning \cmd{git}. Of those, \per{54.27} are declared with the alias name \texttt{gs}, as in \alias{gs}{git status}, and \per{19.19} with the alias name \texttt{gst}, as in \alias{gst}{git status}. All other alias names occur in less than \per{5} of aliases for \texttt{git status}.
		}
    \label{tab:command-summary}
    \newcommand{\numx}[1]{{\footnotesize (\num{#1})}}
\begin{tabular}{@{}l@{}rll@{}}
    \toprule
                &           \% &            Arguments &                                                                 Aliases (\%) \\
    \midrule
     \verb|git| &   \num{5.85} &        \verb|status| &                              \verb|gs| \numx{54.27}, \verb|gst| \numx{19.19} \\
                &   \num{3.48} &              \verb|| &                                \verb|g| \numx{75.71}, \verb|gti| \numx{5.74} \\
                &   \num{3.20} &      \verb|checkout| &      \verb|gco| \numx{50.52}, \verb|gc| \numx{13.87}, \verb|gch| \numx{7.56} \\
                &   \num{3.18} &          \verb|push| &      \verb|gp| \numx{46.73}, \verb|gps| \numx{9.23}, \verb|push| \numx{7.56} \\
                &   \num{3.16} &          \verb|diff| &                                                       \verb|gd| \numx{79.89} \\
                &   \num{2.86} &          \verb|pull| &      \verb|gpl| \numx{18.30}, \verb|gl| \numx{16.59}, \verb|gp| \numx{15.07} \\
                &   \num{2.78} &        \verb|branch| &                               \verb|gb| \numx{73.54}, \verb|gbr| \numx{6.57} \\
                &   \num{2.71} &           \verb|add| &                                                       \verb|ga| \numx{80.96} \\
                &   \num{2.00} &        \verb|commit| &                               \verb|gc| \numx{63.16}, \verb|gci| \numx{5.33} \\
                &   \num{1.96} &     \verb|commit -m| &       \verb|gcm| \numx{31.29}, \verb|gc| \numx{25.18}, \verb|gm| \numx{7.97} \\
    \midrule
      \verb|ls| &  \num{14.45} &  \verb|--color=auto| &                                                       \verb|ls| \numx{99.04} \\
                &   \num{8.63} &            \verb|-A| &                                                       \verb|la| \numx{97.61} \\
                &   \num{7.80} &           \verb|-CF| &                                                        \verb|l| \numx{98.75} \\
                &   \num{6.78} &          \verb|-alF| &                                                       \verb|ll| \numx{97.49} \\
                &   \num{5.46} &            \verb|-l| &                                 \verb|ll| \numx{78.83}, \verb|l| \numx{7.91} \\
                &   \num{3.75} &              \verb|| &                                \verb|l| \numx{27.90}, \verb|sl| \numx{21.45} \\
                &   \num{2.88} &            \verb|-G| &                                                       \verb|ls| \numx{96.47} \\
                &   \num{2.74} &           \verb|-la| &      \verb|ll| \numx{38.42}, \verb|la| \numx{26.87}, \verb|lla| \numx{12.63} \\
                &   \num{2.67} &            \verb|-a| &                                                       \verb|la| \numx{76.94} \\
                &   \num{1.92} &           \verb|-al| &         \verb|ll| \numx{49.69}, \verb|la| \numx{12.23}, \verb|l| \numx{8.49} \\
    \bottomrule
\end{tabular}%

\end{table}

\subsection{Inductive Coding}

To capture the range of patterns and use cases for which aliases are defined, we analyzed the dataset using inductive coding, a classic technique for qualitative data analysis \citep{saldana:16,thomas:06,dey:03}.
Inductive coding is used when conducting exploratory research without prior expectations on themes in the data.
The individual data points---in our case, alias definitions---are labelled with descriptive tags which try to capture the essence of the datum for later purposes of categorization.
It is an iterative process between theoretical sampling and comparing data within emerging themes, continuing in cycles until no new themes emerge.

Since manually coding the entire dataset is infeasible, we developed our themes by coding a representative sample.
For this sample, we gathered the top three most common aliases for the top ten most common arguments for the top 50 commands (cf.\ Table~\ref{tab:command-summary}), resulting in \num{1381} alias definitions, directly covering \per{28.77} percent of the dataset.
Additionally, we drew a random sample of \num{200} alias definitions from the long tail of unique aliases.
These are aliases that each occur only once in the entire dataset, making up \per{27.53} of all aliases.
The commands that occur in this long tail are distributed in roughly the same manner as the commands in the whole dataset, the top commands being \cmd{cd}, \cmd{git}, \cmd{ssh}, \cmd{ls}, and \cmd{vim}.
Unique aliases often contain user-specific file system paths (e.g. \alias{gitbash}{source /Users/j/mybin/gitsh}), happen to have a unique combination of arguments (e.g. \alias{ls}{ls -GphF}) or are otherwise highly particular (e.g. \alias{h23}{history -23000}).

In total, we looked at \num{1581} aliases during the coding process.
In order to reason about the intent of any particular alias, we had to take the semantics of each command into account, consulting their \texttt{man} pages and other forms of documentation.\footnote{The website \url{https://explainshell.com} has been an indispensable resource.}
To increase the trustworthiness of our codes, coding was performed independently in parallel by the two authors.
After a first iteration, we compared our labels, consolidating different naming conventions.
In consecutive iterations, we identified ways of formalizing the emerged categories, i.e. constructing automated mechanisms for classifying alias definitions as belonging to certain categories.
The suitability for mechanical classification was an important factor for the viability of any emerging themes.
The discussion of these formalizations additionally served to establish a better shared understanding.
Ultimately, we reached a saturation point at which further coding and analysis did not lead to further insights.

\section{Customization Practices}

\begin{table}
    \centering
	\caption{Alias types and customization practices}
    \label{tab:practices}
    \begin{tabular}{llrr}
    %\toprule
    & & \# & \% \\
    \toprule
    \multicolumn{2}{l}{\textsc{Shortcuts}} & & \\
    & Nicknaming Commands       & \num{244872} & 11.11 \\
    & Abbreviating Subcommands  & \num{194850} &  8.84 \\
    & Bookmarking Locations     & \num{321546} & 14.59 \\
    \midrule
    \multicolumn{2}{l}{\textsc{Modifications}} & & \\
    & Substituting Commands     & \num{100564} &  4.56 \\
    & Overriding Defaults       & \num{319239} & 14.48 \\
    & Colorizing Output         & \num{182623} &  8.29 \\
    & Elevating Privilege       &  \num{93683} &  4.25 \\
    \midrule
    \multicolumn{2}{l}{\textsc{Scripts}} & & \\
    & Transforming Data         &  \num{74719} &  3.39 \\
    & Chaining Subcommands      &  \num{22062} &  1.00 \\
    \bottomrule
\end{tabular}
\end{table}

We identified nine customization practices among three types of aliases:
\textsc{Shortcuts} introduce new names and are often used for \emph{nicknaming commands}, \emph{abbreviating subcommands}, and \emph{bookmarking locations};
\textsc{Modifications} change the semantics of commands by \emph{substituting commands}, \emph{overriding defaults}, \emph{colorizing output}, and \emph{elevating privilege};
and \textsc{Scripts} combine multiple commands, often for the purposes of \emph{transforming data} or \emph{chaining subcommands}.
We developed automated classification methods for each practice, which can be found in our replication package.
Table~\ref{tab:practices} gives a quantitative overview of the prevalence of each of these practices in the dataset.
Any alias can be an expression of multiple customization practices at once, and some practices only occur with certain commands.
Table~\ref{tab:practices-by-command} breaks down the customization practices by command, counting the number of aliases that a command is involved in (including aliases that redefine the command).

We will now discuss the alias types and customization practices in more detail.

\newcommand{\rot}[1]{\makebox[1em][l]{\rotatebox{45}{#1}}}

\newcommand{\full}{$\CIRCLE$}
\newcommand{\half}{$\LEFTcircle$}
\newcommand{\empt}{$\Circle$}

\newcommand{\hist}[1]{\includegraphics[height=1em, width=8em, trim=1em 1em 1em 1em, clip]{compression-#1.pdf}}

\newcommand*{\pie}[1]{\begin{tikzpicture}[scale=0.15]%
    \draw (0,0) circle (1);
    \fill[fill opacity=1,fill=black] (0,0) -- (90:1) arc (90:90-#1*3.6:1) -- cycle;
    \end{tikzpicture}}

\begin{table*}
    \centering
    \caption{Customization practices broken down by command. We present a selection of common commands and for each of the nine customization practices show the percentage of occurrences of the command that happen as part of that customization practice, if it is more than \per{1} of all occurrences of the command. Note that a single command occurrence can be part of multiple customization practices at once. The compression ratio plots are log-log histograms, the red line marks a ratio of~1.}
    \vspace{1em}
    \label{tab:practices-by-command}
\begin{tabular}{llr|ccc|cccc|cc|c}
& & \multicolumn{1}{r}{\#} & \multicolumn{1}{c}{\rot{Nicknaming Commands}} & \multicolumn{1}{c}{\rot{Abbreviating Subcommands}} & \multicolumn{1}{c}{\rot{Bookmarking Locations}} & \multicolumn{1}{c}{\rot{Substituting Commands}} & \multicolumn{1}{c}{\rot{Overriding Defaults}} & \multicolumn{1}{c}{\rot{Colorizing Output}} & \multicolumn{1}{c}{\rot{Elevating Privilege}} & \multicolumn{1}{c}{\rot{Transforming Data}} & \multicolumn{1}{c}{\rot{Chaining Subcommands}} & Compression \\
\midrule
\multicolumn{2}{l}{Version Control} & & & & & & & & & & & \\
&           \texttt{git} &  \num{315841} &          \pie{3.44} &              \pie{36.11} &            \pie{1.84} &                       &                     &        \pie{1.26} &                     &                   &           \pie{3.82} &           \hist{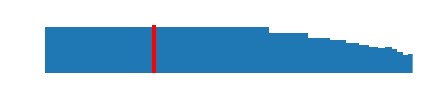} \\
&            \texttt{hg} &    \num{2799} &          \pie{2.47} &              \pie{44.52} &            \pie{2.22} &           \pie{22.15} &                     &        \pie{1.68} &                     &                   &           \pie{3.43} &            \hist{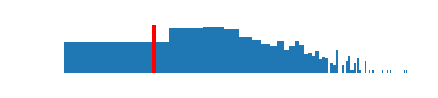} \\
\midrule
\multicolumn{2}{l}{System Tools} & & & & & & & & & & & \\
&            \texttt{ls} &  \num{268423} &           \pie{2.2} &                          &                       &            \pie{3.41} &         \pie{28.08} &       \pie{29.56} &                     &                   &                      &            \hist{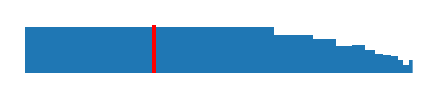} \\
&            \texttt{cd} &  \num{164164} &          \pie{1.07} &                          &           \pie{45.61} &            \pie{1.78} &                     &                   &                     &                   &                      &            \hist{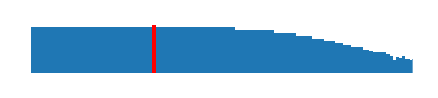} \\
&          \texttt{grep} &  \num{144606} &                     &                          &            \pie{1.86} &            \pie{2.39} &         \pie{67.06} &        \pie{70.6} &                     &       \pie{25.48} &                      &          \hist{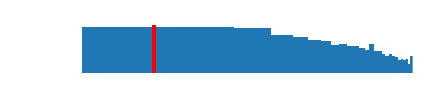} \\
&            \texttt{rm} &   \num{26131} &                     &                          &           \pie{19.67} &             \pie{9.3} &         \pie{49.45} &                   &          \pie{8.43} &                   &                      &            \hist{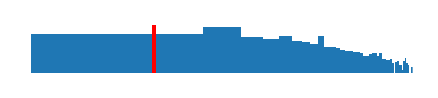} \\
&          \texttt{tmux} &   \num{22821} &           \pie{5.6} &                          &            \pie{4.37} &             \pie{1.7} &         \pie{18.52} &       \pie{26.11} &                     &                   &                      &          \hist{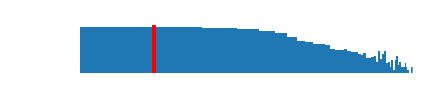} \\
&            \texttt{cp} &   \num{18628} &          \pie{1.14} &                          &           \pie{13.28} &            \pie{3.24} &         \pie{74.37} &                   &           \pie{1.8} &                   &                      &            \hist{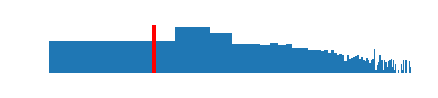} \\
&            \texttt{mv} &   \num{14897} &          \pie{1.91} &                          &             \pie{5.1} &            \pie{3.13} &         \pie{81.23} &                   &          \pie{1.01} &                   &                      &            \hist{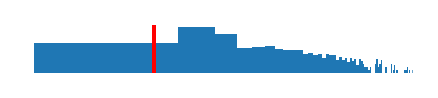} \\
&            \texttt{du} &   \num{12480} &                     &                          &            \pie{3.13} &            \pie{3.91} &         \pie{45.45} &        \pie{2.01} &          \pie{1.75} &                   &                      &            \hist{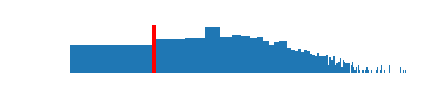} \\
&          \texttt{sort} &   \num{10802} &                     &                          &                       &            \pie{1.46} &                     &                   &                     &       \pie{95.97} &                      &          \hist{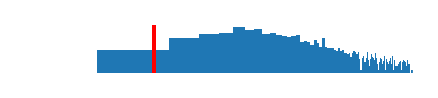} \\
&         \texttt{mkdir} &   \num{10351} &         \pie{10.11} &                          &            \pie{4.37} &            \pie{2.02} &         \pie{57.72} &                   &                     &                   &                      &         \hist{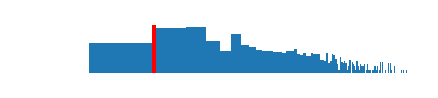} \\
&            \texttt{df} &   \num{10266} &                     &                          &                       &            \pie{4.41} &         \pie{82.09} &                   &                     &                   &                      &            \hist{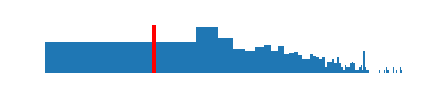} \\
&          \texttt{diff} &    \num{4697} &                     &                          &                       &           \pie{46.18} &         \pie{32.13} &       \pie{91.95} &                     &                   &                      &          \hist{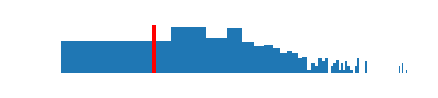} \\
\midrule
\multicolumn{2}{l}{Text Editors} & & & & & & & & & & & \\
&           \texttt{vim} &   \num{99521} &          \pie{8.68} &                          &           \pie{29.12} &           \pie{50.52} &          \pie{1.68} &                   &          \pie{3.41} &                   &                      &           \hist{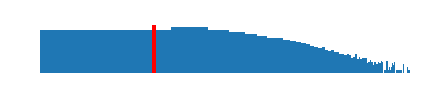} \\
&         \texttt{emacs} &   \num{12990} &         \pie{11.74} &                          &            \pie{8.31} &           \pie{18.41} &         \pie{12.04} &         \pie{1.8} &          \pie{2.03} &                   &                      &         \hist{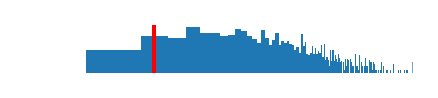} \\
&           \texttt{sed} &    \num{7545} &                     &                          &           \pie{20.32} &           \pie{14.04} &          \pie{2.03} &        \pie{2.68} &                     &       \pie{70.01} &                      &           \hist{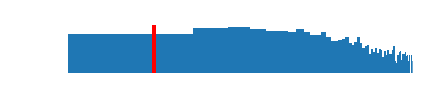} \\
&          \texttt{subl} &    \num{5030} &          \pie{10.6} &                          &           \pie{31.49} &            \pie{44.0} &          \pie{1.39} &                   &          \pie{1.93} &        \pie{1.15} &                      &          \hist{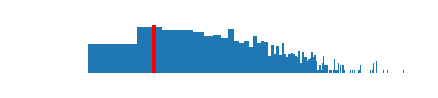} \\
&          \texttt{nano} &    \num{4030} &          \pie{6.67} &                          &           \pie{36.97} &           \pie{11.17} &         \pie{16.63} &                   &         \pie{16.72} &                   &                      &          \hist{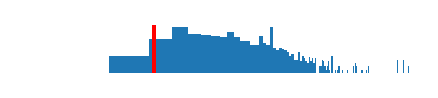} \\
\midrule
\multicolumn{2}{l}{Infrastructure} & & & & & & & & & & & \\
&        \texttt{docker} &   \num{39111} &         \pie{16.36} &              \pie{23.87} &           \pie{11.48} &                       &                     &                   &          \pie{4.19} &                   &           \pie{2.75} &        \hist{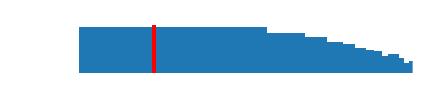} \\
&       \texttt{kubectl} &   \num{12610} &         \pie{20.37} &              \pie{13.94} &            \pie{2.31} &                       &          \pie{1.03} &                   &                     &                   &                      &       \hist{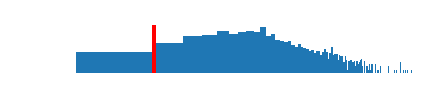} \\
&       \texttt{vagrant} &    \num{6847} &         \pie{10.79} &              \pie{63.27} &            \pie{1.04} &                       &                     &                   &                     &                   &            \pie{4.7} &       \hist{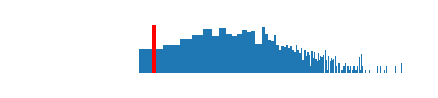} \\
\midrule
\multicolumn{2}{l}{Networking} & & & & & & & & & & & \\
&           \texttt{ssh} &   \num{32573} &                     &                          &           \pie{65.43} &             \pie{1.9} &          \pie{3.86} &        \pie{1.95} &                     &                   &                      &           \hist{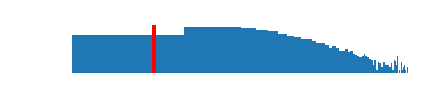} \\
&          \texttt{curl} &   \num{10558} &                     &                          &           \pie{80.24} &            \pie{4.23} &          \pie{3.06} &                   &                     &                   &                      &          \hist{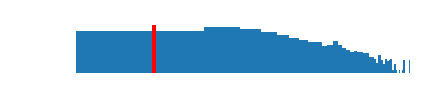} \\
&          \texttt{wget} &    \num{3937} &           \pie{1.3} &                          &           \pie{28.63} &            \pie{7.47} &         \pie{38.96} &                   &                     &                   &                      &          \hist{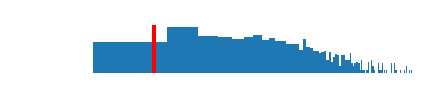} \\
\midrule
\multicolumn{2}{l}{Package Managers} & & & & & & & & & & & \\
&          \texttt{apt} &    \num{17632} &          \pie{1.28} &              \pie{58.35} &                       &            \pie{3.06} &          \pie{4.37} &                   &         \pie{75.24} &                   &          \pie{17.04} &           \hist{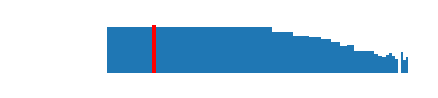} \\
&        \texttt{pacman} &   \num{14798} &          \pie{1.33} &              \pie{67.02} &                       &            \pie{4.09} &          \pie{7.57} &        \pie{7.12} &         \pie{63.02} &                   &           \pie{1.19} &        \hist{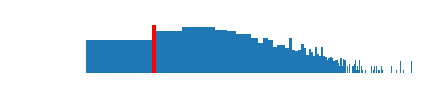} \\
&          \texttt{brew} &    \num{8555} &          \pie{3.12} &              \pie{32.24} &                       &            \pie{1.48} &                     &                   &          \pie{1.68} &                   &          \pie{33.77} &          \hist{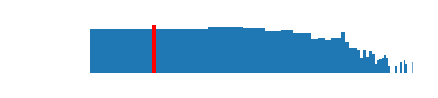} \\
\bottomrule
\end{tabular}
\end{table*}

\subsection{Shortcuts}

\begin{figure}
	\centering
	\includegraphics[width=\columnwidth]{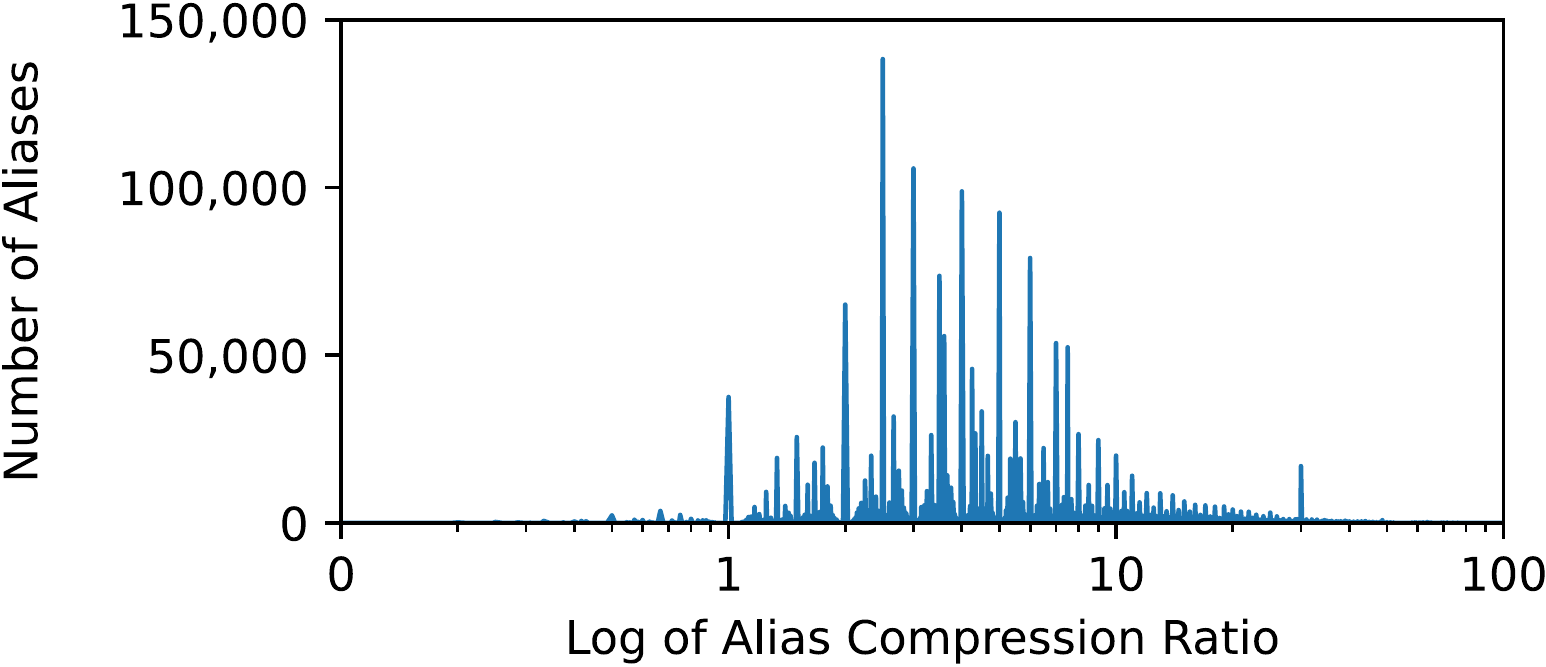}
	\caption{Distribution of alias compression ratios}
	\label{fig:compression}
\end{figure}

The most obvious use of an alias is to give a complex expression a short and/or memorable name.
The average length of an alias name is 4.3~characters, whereas the average length of an alias value is 23.7~characters.
If we divide the length of an alias value by the length of the alias name, we get the \emph{compression ratio} of the alias.
For example, the alias \alias{gs}{git status} has a compression ratio of~5.
Fig.~\ref{fig:compression} shows the distribution of compression ratios over all aliases in the dataset.
The median compression ratio is 4.25, meaning half of all alias values are at least four times as long as their alias names.
A compression ratio less than 1 indicates a name that is longer than the value it aliases.

There are \num{26055} aliases (\per{1.18}) with names longer than their values.
The two longest alias names we found are from joke definitions.
The first is \num{1772} characters long and is comprised of the letter `f' repeated \num{1053} times, followed by the letter `u' repeated 719 times.
It is an alias for the \verb|cat| command with a similarly named file as an argument.
The second longest alias name is a Swedish compound word of \num{131} characters,\footnote{Translating, roughly, to northwestern-glacier-artillery-flight-thrust-simulator-plant-equipment-maintenance-follow-up-systems-discussion-posts-preparation-works.} aliasing the \verb|ls| command.

On the other end of the spectrum, an alias named \texttt{line} echoes \num{23635} dashes, achieving a compression ratio of \num{5911}, the highest among all aliases.
The second highest comes from an alias named \verb|BEEP|, which invokes the Linux \verb|beep| utility 9 times in succession, with a combined \num{4471} arguments.
When executed, it appears to play Daft Punk's 2001 instrumental single \emph{Aerodynamic}.

Beyond just compression and expansion of strings, we can see a few distinct customization practices related to naming.

\paragraph{\bf Nicknaming Commands.}

There are \num{244872} aliases in our data\-set (\per{11.11}) that merely give a new name to a command, without adding any arguments, and without the name belonging to a different command (that would be a substitution, see below).
The most often occurring nicknames are \alias{g}{git}, \alias{c}{clear}, \alias{h}{history}, and \alias{v}{vim}.
Almost all (\per{93.03}) of these kinds of aliases introduce a nickname that is shorter than the command they are referring to, and about half (\per{50.58}) introduce a name that is only one or two characters long.

A special case of nicknaming occurs when the new name is a common misspelling of the command.
In this case, the alias acts like an autocorrect mechanism, as in \alias{got}{git}.
%While it is easy for the human eye to determine instances of these typographical errors, it is not as straightforward to formalize all different cases and to distinguish them from regular command shortcuts.
%We opt for a conservative criterion (potentially underestimating the true extent of the phenomenon) that looks only at aliases whose names are of the same length as their aliased commands, with a string distance measure above an empirically determined threshold.
To determine instances of these typographical errors, we surveyed and experimented with different string distance measures~\citep{navarro:01} and decided on using the Damerau-Levenshtein algorithm~\citep{damerau:64}.
%It is a robust measure that in addition to tracking the number of insertions, deletions, and substitutions between two strings, also captures the transposition of two characters, a common occurrence in misspelled commands.
%We computed the distance measure for all applicable aliases, and 
We determined empirically that a distance measure of 2 seems like a good threshold to decide whether or not an alias corrects a misspelling.
We found \num{9195} aliases (\per{0.42}) that serve as autocorrect rules, most commonly involving transposition (\alias{grpe}{grep}), case-sensitivity (\alias{Jupyter}{jupyter}), localization (\alias{pluralise}{pluralize}), and punctuation (\alias{docker-build}{docker\_build}).

%On the flip side, aliases are also used to disable autocorrect mechanisms.
%The Z shell has built-in spelling correction, which can be selectively disabled using the \verb|nocorrect| command.
%There are \num{7326} aliases (\per{0.33}) in our dataset disabling Z shell autocorrection for certain commands, most commonly the filesystem commands \verb|mv|, \verb|mkdir|, \verb|cp| and \verb|rm|.

\paragraph{\bf Abbreviating Subcommands.}

Many commands can operate in different modes, or act as interfaces to a variety of different \emph{subcommands}.
The subcommand is commonly specified as the first argument to the command, and takes its own set of arguments and flags.
For example, \texttt{git push --tags} executes the \texttt{push} subcommand of \cmd{git} with the \texttt{--tags} flag enabled.
We identified 67 commands in our dataset that take subcommands, such as \cmd{git}, \cmd{docker}, or \cmd{systemctl}.
Noticeably, we found \num{194850} aliases (\per{8.84}) that are purely abbreviations of subcommands, without adding any additional arguments beyond the subcommand.
For example, \alias{gs}{git~status} or \alias{gd}{git diff}.
The majority of such subcommand abbreviations (\per{58.5}) are for \cmd{git}, with \num{113980} aliases defined purely for abbreviating \cmd{git} subcommands, accounting for \per{36.77} of all aliases involving \cmd{git}.
The command with the second-most subcommand abbreviations is the package manager \cmd{pacman}, with only \num{9918} instances (\per{5.09} of subcommand abbreviations, but \per{68.67} of all aliases involving \cmd{pacman}).

\paragraph{\bf Bookmarking Locations.}

When an aliased command is called with an argument that references some specific local or remote location, like a file path or domain, the alias acts as a bookmark to that location.
For instance, \alias{dl}{cd \textasciitilde/Downloads} and \alias{starwars}{telnet towel.blinkenlights.nl} are both bookmark aliases.
To find such bookmarking uses in our dataset, we searched for arguments that are locations, which we take to be any of the following:
\begin{itemize}
    \item A string containing a forward slash (\verb|/|), indicating a path.
    \item An IPv4 address, matched by the liberal regular expression\\\verb|[0-9]+\.[0-9]+\.[0-9]+\.[0-9]+|
    \item A string containing one of the known top-level domains\footnote{\url{http://data.iana.org/TLD/tlds-alpha-by-domain.txt}} preceded by a dot (\verb|.|) and followed by a slash (\verb|/|), colon (\verb|:|) or the end of the string.
\end{itemize}
To avoid false positives, we sampled the top 300 search results according to the above criteria and determined some exclusion patterns.
For instance, \texttt{/dev/null} is not a location for our purposes.
Neither is \texttt{origin/master}, and thus an alias like \alias{gm}{git~merge~origin/master} does not count as a bookmark.
We also exclude aliases that are merely referencing unnamed relative directories (e.g., \verb|../..|).

By our definition, \num{321546} aliases (\per{14.59}) are bookmarks.
Of these, \num{59931} are remote bookmarks containing URLs or IP addresses (\per{15.92} of all bookmarks).
Bookmarks are used predominantly for file system navigation, and the \verb|cd| command is featured heavily.
Most other uses seem to be development related, like starting services such as web servers or databases with pre-defined locations, opening frequently edited files, or outputting logs, as in \alias{onoz}{cat /var/log/errors.log}
%, or opening frequently edited files, as in the most common alias for the most common location, which is for editing the shell configuration itself: \alias{zshconfig}{vim \textasciitilde/.zshrc}.
%Table~\ref{tab:locations} shows the top local and remote locations found in aliases.

% \begin{table}
%     \caption{Top 5 local and remote locations found in aliases}
%     \label{tab:locations}
%     \input{tables-locations.tex}
% \end{table}

\subsection{Modifications}

Aliases are not only used syntactically, for naming purposes, but also in ways that change the semantics of certain commands.
We found four customization practices related to command modification.

\paragraph{\bf Substituting Commands.}

When an alias name is identical to the name of a pre-existing command, the alias defines a substitution for that command.
A common example is \alias{more}{less}, replacing a standard Unix utility (\cmd{more}) with a more capable but similar command (\cmd{less}).
This can also be used for subterfuge, as in \alias{emacs}{vim} (appearing 132 times in our dataset) or indeed \alias{vim}{emacs} (86 times, alas).

To determine which alias names are also actual command names, we compared them to known Unix commands\footnote{\url{https://en.wikipedia.org/wiki/List_of_Unix_commands}\\ and \url{https://en.wikipedia.org/wiki/List_of_GNU_Core_Utilities_commands}} and a curated sample of commands from our dataset (taking care to not include names that appear in a command position but are actually just other aliases).
To determine proper substitutions, we only count aliases whose value does not also include the name of the command (which would point to an overriding alias, see below).
We find that \num{100564}~aliases (\per{4.56}) are used to substitute one command for another.
The top three substitutions are \verb|vi| $\rightarrow$ \verb|vim|, \verb|vim| $\rightarrow$ \verb|nvim|, and \verb|vi| $\rightarrow$ \verb|nvim|.

\paragraph{\bf Overriding Defaults.}

When an alias has the same name as the command it aliases, as in \alias{ls}{ls -G}, then the alias re-defines the command and effectively overrides its default settings.
Any time the command is now executed, it will be with the arguments specified in the alias.
There are \num{319239} aliases in our dataset (\per{14.48}) that are used to override defaults in this way.
Aliases to override the defaults of the \cmd{grep} family of commands (\cmd{grep}, \cmd{egrep}, \cmd{fgrep}) occur \num{96970}~times, accounting for \per{4.4} of all alias definitions (and \per{68.27} of all \cmd{grep} appearances).
The \cmd{ls} command is redefined with new defaults \num{75374} times, accounting for \per{3.42} of all aliases (\per{28.99} of \cmd{ls} appearances).

Looking at the new defaults of these redefined commands, they reveal a variety of user preferences, especially in the diverse long tail, where we find a lot of unique alias definitions and argument combinations.
Two areas of customization stand out, however: formatting output and adding safety.
The majority of overrides for file system commands (\cmd{mv}, \cmd{cp}, and \cmd{rm}, but also \cmd{ln}, for creating symbolic links) enable interactive mode (\texttt{-i} and variations), which prompts the user before performing potentially destructive actions.
Verbose output (\texttt{-v}) also plays a role here, describing exactly what kind of effects a command execution had or will have.
Enabling verbosity can also be seen as a kind of output formatting, although much more common is the wish for human-readable output.
For example, the alias \alias{df}{df -h} ensures that the available disk space is displayed in common size units, as opposed to just the raw number of bytes.
But by far the most common reason for overriding defaults is to enable colorized output.
This behavior is so prevalent that we count it as a customization practice in its own right.

\paragraph{\bf Colorizing Output.}

Enabling colored output can be done in many different ways: adding an argument (like \texttt{less -R} or \cmd{grep --color=always}), setting an environment variable (as in \alias{ssh}{TERM=xterm256color ssh}), running the command through a tool that colorizes its output (like \cmd{grcat} or \cmd{pygmentize}), or even replacing a command outright (\alias{diff}{colordiff}).
Taking all these varieties into account, more than half of all command redefinitions (\per{57.21}) enable colored output by default.
This amounts to a surprising \num{182623} aliases, or \per{8.29} percent of all aliases in the dataset.
If we extend this count to also include aliases that introduce new names (like \alias{ll}{ls -l --color=auto}), then more than \per{10} of aliases colorize a command's output.

\begin{figure*}
	\centering    
	\includegraphics[width=0.73\linewidth]{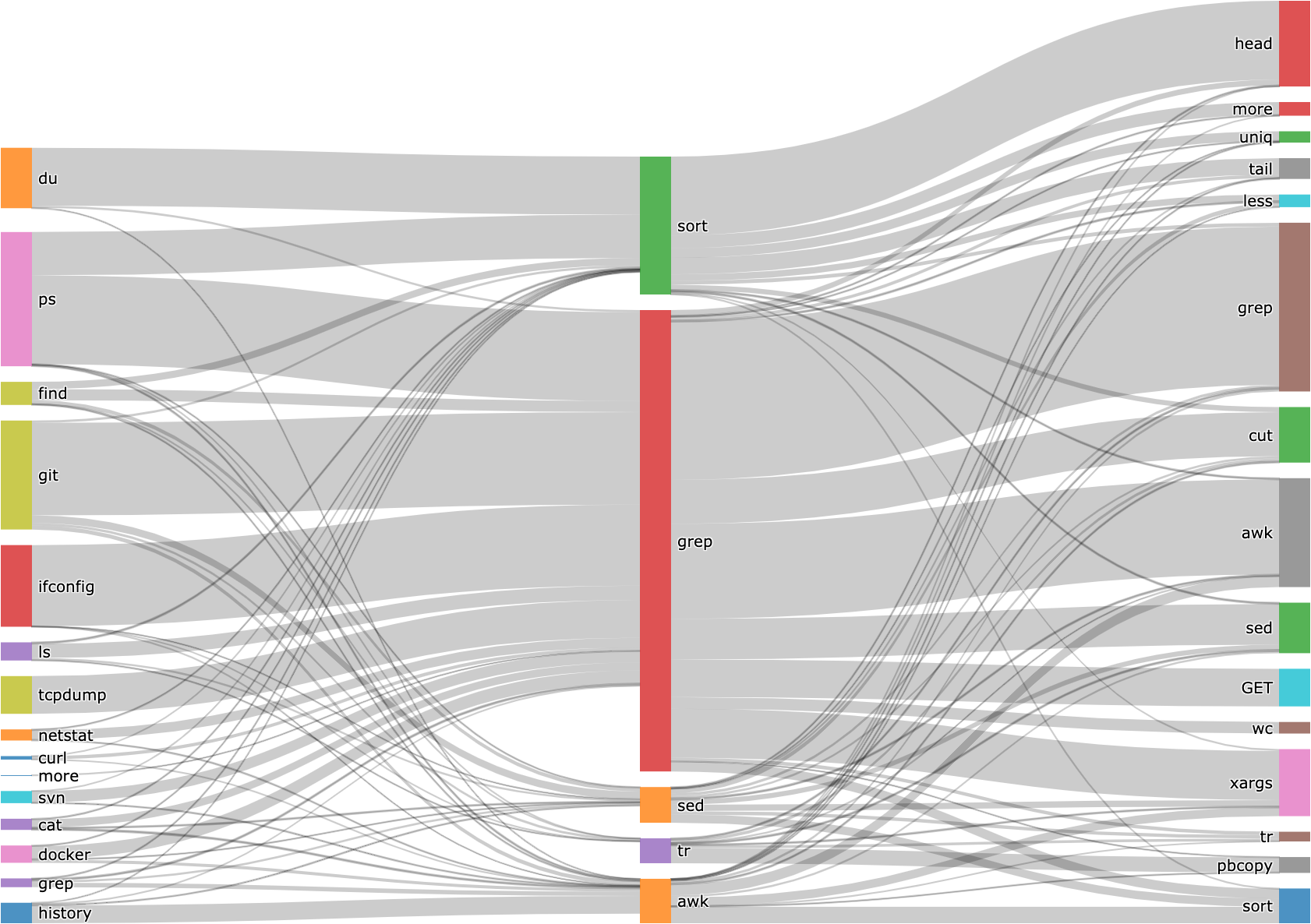}
	\caption{Flow diagram of the top 250 pipelines with three commands that make up at least \per{10} of one command's usage}
	\label{fig:flow}
\end{figure*}

\paragraph{\bf Elevating Privilege.}

The \cmd{sudo} command allows the user to execute another command with superuser privileges.
Combining a command with \cmd{sudo} is often necessary if the other command needs to modify critical parts of the system.
In our dataset, we found \num{93683} aliases (\per{4.25}) in which a command is prefixed with \cmd{sudo}.
The top \cmd{sudo}-prefixed command is the package manager \cmd{apt-get}, appearing \num{10467} times with \cmd{sudo}.
Remarkably, these are \per{89.35} of all occurrences of \cmd{apt-get}.
In fact, \per{72.45} of all occurrences of the package managers \cmd{apt}* (Debian and derivatives; including \cmd{apt}, \cmd{apt-get}, \cmd{apt-cache}, \cmd{aptitude}, and \cmd{\$apt\_pref}), \cmd{pacman}, \cmd{abs} and \cmd{aur} (Arch Linux), \cmd{yum} (RPM), \cmd{dnf} (Fedora), \cmd{zypper} (openSUSE), \cmd{port} (macOS), and \cmd{gem} (Ruby) are together with \cmd{sudo}, and these package managers account for \per{29.1} of all \cmd{sudo} occurrences.
Interestingly, the macOS package manager \cmd{brew} rarely appears with \cmd{sudo} (only \per{1.07}), even though it is the third most occurring package manager overall, behind \cmd{apt}* and \cmd{pacman}.

Other commands that more often than not demand elevated privileges are system utilities like \cmd{systemctl}, \cmd{shutdown}, \cmd{lsof} or \cmd{mount}.

\subsection{Scripts}

Aliases that combine multiple commands are basically tiny shell scripts.
In our dataset, \num{204142} aliases (\per{9.26}) compose multiple commands.
The most popular composition operator is the pipe (\verb`|`), used in \per{39.66} percent of alias scripts, followed by the operators for simple chaining (\verb|;|), with \per{29.61}, and logical conjunction (\verb|&&|), with \per{26.88}.
Other operators (\verb`||`, \verb`|&`) appear in only \per{3.85} of multi-command aliases.

There are two scripting practices that are of particular interest.

\paragraph{\bf Transforming Data.}

The pipe (\verb`|`) creates an interface between two otherwise separate programs.
It embodies the Unix philosophy of small tools doing one thing well, which can then be connected together to accomplish more complex tasks.
There are \num{74719} aliases (\per{3.39}) combining two or more commands using only the pipe operator.
The most common command occurring after a pipe, by far, is \cmd{grep}, which makes an appearance in almost half of all pipelines (\per{46.16}), more than three times as often as \cmd{xargs} and \cmd{sort}.
The most common data sources are \cmd{ps}, \cmd{git}, and \cmd{ls}, which are found at the beginning of almost a third (\per{32}) of all pipelines.
Fig.~\ref{fig:flow} shows a flow diagram of the top pipelines with three commands.

The names of aliases for such pipelines are varied, speaking to the broad range of tasks that can be accomplished by combining various Unix tools. They range from the descriptive, as in \alias{diskspace}{du~-S~|~sort~-n~-r~|~more} or  \alias{weather}{wget~-qO~-~http://wttr.in/~|~head~-7}, to the very terse, as in \alias{h}{history~|~uniq~|~tail~-15} or \alias{lll}{ls~-trlh~|~less}.
Interestingly, aliases with the same name usually describe pipelines with the same general shape (the same commands in the same order), but slightly different argument combinations:
\begin{Verbatim}[commandchars=\\\{\},codes={\catcode`$=3\catcode`^=7}]
lsd $\rightarrow$ ls -l | grep "\^{}d"
lsd $\rightarrow$ ls -la | grep \^{}d
lsd $\rightarrow$ ls -lGFA --color | grep -i "\^{}d.*/"
lsd $\rightarrow$ ls -lh | grep --color=never '\^{}d'
\end{Verbatim}
% or
% \begin{Verbatim}[commandchars=\\\{\},codes={\catcode`$=3\catcode`^=7}]
% pg $\rightarrow$ ps aux | grep
% pg $\rightarrow$ ps -Af | grep \$1
% pg $\rightarrow$ ps -ef | grep
% pg $\rightarrow$ ps aux | grep -i
% \end{Verbatim}
% and so on.
This highlights the highly personal nature of aliases, each customized for an individual use case.

\paragraph{\bf Chaining Subcommands.}

An interesting pattern appearing in alias scripts are chains of subcommand invocations.
For example, the package manager \cmd{brew} has a subcommand \texttt{update}, for updating the package database, and a subcommand \texttt{upgrade}, for upgrading previously installed packages to the latest available versions.
\per{28.08} of all aliases involving the \cmd{brew} command contain the composition \verb|brew update && brew upgrade| (sometimes with \verb|;| instead of \verb|&&|), with alias names like \verb|update|, \verb|brewup|, \verb|bup|, etc.
This pattern of repeated subcommand invocations can be found in \num{22062} aliases (\per{1}), and it is most prevalent among package managers, like \cmd{brew}, \cmd{apt-get}, \cmd{npm} or \cmd{gem}, mostly for the same purpose as above.

The command with the highest absolute number of aliases showing this pattern is \cmd{git}, however, with \num{12063} occurrences (\per{3.89} of all aliases using \cmd{git}).
Here, the uses are more varied, 
e.g., \alias{commit}{git add . \&\& git commit -m}, 
or \alias{gitpull}{git stash \&\& git pull \&\& git stash pop},
or indeed \alias{whoops}{git reset --hard \&\& git clean -df}.

\section{Implications}

Through our large-scale analysis of the collective knowledge of shell customization via aliases,
%of unique files that contain over 2.2 million alias definitions, 
we gained insight into practices detailing how users customize their command-line interface.
Based on our observations, we outline discussion points that go beyond single customization practices and identify implications that can address shortcomings in command-line usability and tie them to existing user experience research.
Further, while our presented findings already give us an understanding of customization practices over many different kinds of commands, we view our collected dataset as a playground for fine-grained discovery that can benefit researchers, tool builders, and command-line users.

\subsection{Learning Repair Rules}
\label{sec:repair-rules}

The complexity of commands and arguments can cause users to introduce errors when working in a command-line interface.
Figuring out specifically how to fix these errors is often a convoluted process.
A popular open source project that attempts to navigate this issue\footnote{\url{https://github.com/nvbn/thefuck}} uses a set of rules to suggest possible error corrections for commands.
While these rules are all hard-coded, we envision leveraging the global wisdom of customizations in our large-scale dataset to learn rules that form the basis for different kinds of suggestions.
This is in line with visions of integrating collective intelligence in software development \citep{bruch:10}, in particular work in leveraging emergent behavior from corpora~\citep{fast:14} that we can codify based on our customization data.
We can also see approaches similar to work on learning code completions from examples \citep{bruch:09}, with our dataset of alias definitions serving as an oracle for an automatic software repair system \citep{monperrus:18} in the domain of shell commands.
Using our dataset of known-good command invocations, it should be possible to train a statistical language model for command repair, akin to related work in code synthesis~\citep{raychev2014completion}.

As an example, take the following erroneous invocation:
\begin{Verbatim}[breaklines=true]
$ apt-get install vim
E: Could not open lock file /var/lib/dpkg/lock - open (13: Permission denied)
E: Unable to lock the administration directory (/var/lib/dpkg/), are you root?
\end{Verbatim}
Without having to consult a hard-coded rule involving knowledge about \cmd{apt-get}, or even looking at the specific error that is produced, a command repair system trained on our dataset of alias definitions could easily suggest the correct fix: \texttt{sudo apt-get install vim}.
It is reasonable to assume that this could be inferred as the correct invocation, because in aliases the command sequence \texttt{apt-get install} occurs almost exclusively pre-fixed with \cmd{sudo}.

As another example, the following error is caused by the wrong order of arguments to the \cmd{systemctl} command:
\begin{Verbatim}
$ systemctl docker status
Unknown command verb docker.
\end{Verbatim}
The correct invocation is \texttt{systemctl status docker}.
It is again very plausible that a repair rule for this type of error could be learned from our dataset, based on the prevalence of aliases containing the command \cmd{systemctl} together with an argument \texttt{status} that occurs in first position,
%\footnote{There are \num{652} such aliases in our dataset, out of \num{709} aliases involving \cmd{systemctl} and a \texttt{status} argument.}
indicating the latent knowledge that \texttt{status} is in fact a subcommand of \cmd{systemctl}.

\subsection{Discovering Workflows}
\label{sec:discovering-workflows}

Following a different thread of leveraging emergent practices, we can also see how our dataset would enable a world beyond only trying to fix immediate errors, by providing usage hints that could introduce users to common parameters and workflows.
For example, as soon as a user tries to \cmd{sort} the output of the \cmd{ps} command, the alias \alias{mem10}{ps auxf | sort -nr -k 4 | head -10} can serve as a suggestion for the complex but common data transformation that results in showing the ten most memory-intensive processes.
%\paragraph*{\bf Object Protocols}
Similarly, in the practice of \emph{chaining subcommands} we can clearly see the prevalence of object protocols~\citep{beckman:11}, which are implicit rules determining the order in which commands have to be executed.
We can improve usability by enabling the discovery of these implicit rules and by exposing the dependency structure based on our customization data.
For instance, if executing \verb|brew upgrade| results in a failure, we can suggest using \verb|brew update && brew upgrade| instead, based on the patterns in our dataset (cf. Section~\ref{sec:repair-rules}).

Our findings can also contribute to recent work on the parallelization and distribution of shell scripts.
Systems like PaSh \citep{pash} and POSH \citep{posh} rely on manual annotation of commands and their arguments to effectively parallelize shell scripts.
Our data can help focus these annotation efforts by informing the developers of these systems about which groups of commands and arguments are most frequently used together.
The KumQuat system \citep{kumquat} leverages program synthesis techniques to search a large space of candidate solution to synthesize parallel shell scripts.
The collective knowledge present in alias definitions can guide this search and justify certain intuitions about the latent data parallelism in Unix pipelines \citep{odfg}.
For example, while a parallel version of the \cmd{comm} command for comparing sorted files line-by-line is not synthesizable in general, it becomes trivially parallelizable if each of its input lines is known to be unique.
Evidence that this indeed the common case can be found in our dataset, where \per{41.29} of all occurrences of \cmd{comm} follow \texttt{sort~|~uniq} or \texttt{sort~-u}, and the remainder mostly have unique data sources as input, like \texttt{pacman -Qeq}.

\subsection{Uncovering Conceptual Design Flaws}

Customization can also be an indicator for problems in the underlying conceptual design, manifesting as usability frustrations that require adaptation by the user.
In their analysis of Git, \citet{perez:13,perez:16} describe a number of flaws and operational misfits arising from the conceptual design of the software.
The frustrations experienced by users because of these design flaws are evident based on the alias definitions in our dataset.

For example, the difficulties some Git users have with the concept of \emph{staging} can be seen in aliases that ensure untracked files are included in a commit by explicitly adding them beforehand, like \alias{commit}{git add . \&\& git commit -m} or \alias{gac}{git add --all \&\& git commit}.\footnote{See ``Just Let Me Commit!'' in \cite{perez:13} and ``Incoherent Commit'' in \cite{perez:16}. Confusingly, \texttt{git commit -a}, while performing an implicit \texttt{add}, does not include untracked files.} Another frustration is having to use \texttt{git stash} to temporarily save uncommitted changes and clean the working directory in order to avoid conflicts when using other Git commands.
Stashing in itself has no higher purpose in version control, it merely exists as a concept to work around limitations in Git.\footnote{See ``I Just Want To Switch Branches!'' in \cite{perez:13}, and especially ``Unmotivated Stash'' and ``Branch Coupling'' in \cite{perez:16}.}
This can be seen in aliases like \alias{gspull}{git stash \&\& git pull \&\& git stash pop}, which defines a new type of pull command that stashes away ongoing work before pulling in remote changes and finally re-applying the stashed work.
The same problem happens when switching branches, hence aliases like \alias{gsc}{git stash \&\& git checkout \$1 \&\& git stash pop}.

\citet{church:14} found that version control systems are generally perceived as being risky to use, and sought explanations for this impression via an analysis of Git using a framework of cognitive dimensions \citep{green:96}.
One of the dimensions that dominate the command-line interface of Git is \emph{Hidden Dependencies}.
The are many hidden dependencies in Git, a prominent one being the dependency between the local branch and the remote repository.
This is revealed by alias definitions like \alias{gitstatus}{git~remote~update \&\& git~status}.
Unless one first manually updates Git's local information about remote branches, the command \texttt{git status} will happily report that the local branch is up-to-date with respect to its remote origin, even if the remote repository is in fact many commits ahead.

We want to emphasize that we are not suggesting that large-scale quantitative data of customization practices can replace qualitative analysis, but rather that the corpus we provide, together with our findings, can support exploration and provide new insights for usability research.
Alias definitions can provide evidence for analytic theories based on cognitive or conceptual models of software use, because they codify workarounds for common annoyances and other customizations based in every-day use.
According to a recent need-finding study by \citet{zhang:20}, API designers have a strong desire to know more about users' mental models, and wish to validate design hypotheses with examples of real-world API usage.
Existing techniques for mining API usage fall short in this respect, and the study highlights the importance of, among other things, looking at how users deal with unanticipated corner cases and how they apply workarounds.
We suspect makers of command-line software are in a similar situation as API designers and could similarly benefit from community usage data that highlights gaps between interface design and users' expectations.

\subsection{Contextual Defaults}
\label{sec:contextual-defaults}

Choosing proper defaults in user interfaces is a pillar of user experience design~\citep{nielsen2005power}.
The fact that \per{14.48} of the customizations in our dataset are for \emph{overriding defaults} suggests that, at least for some groups of users, the default settings of their tools could be improved.
We see \emph{overriding defaults} not necessarily as an indictment of the involved commands, but rather as an indication that the assumed user context does not in all cases match the actual usage profile.
This can be the case if the tool assumes a different execution environment than the one it is ultimately used in, e.g. personal notebook vs cloud deployment (where an alias like \alias{java}{java -ea -server} ensures that Java programs are always run on a server-optimized virtual machine) or interactive terminal vs shell script use (cf. Section~\ref{sec:interactivity-vs-scripting}), or if the tool assumes a certain type of user with different needs than the actual user.

Indeed, the variety of different defaults in the data indicate what we call \emph{contextual} defaults, where context could be a reflection of the level of expertise of a command-line user, or a certain persona (e.g., system administrator, data scientist, or software engineer).
For example, the top default alias for the \cmd{ffmpeg} command is \alias{ffmpeg}{ffmpeg -hide\_banner}, suppressing verbose default output that can be confusing for newcomers but is helpful for the tool developers when providing support and locating errors.\footnote{Coincidentally, we also found a ticket in the project's issue tracker requesting this top most default argument from our dataset to become the default option for the command: \url{https://trac.ffmpeg.org/ticket/7211}}
We could imagine providing different sets of defaults to different users, effectively alias starter packs, generated from our data.
We see parallels to work that investigates contextual preferences and personalization in information systems~\citep{de:15, stefanidis:11} and privacy research~\citep{wijesekera:18, alom:19}.

\subsection{Interactivity vs Scripting}
\label{sec:interactivity-vs-scripting}

The first ``modern'' command line, the Bourne shell from 1977, had two primary goals: to provide an interactive command interpreter, and at the same time serve as a scripting system \citep{jones:11}.
There is a natural tension between these two goals, which becomes evident when users are \emph{overriding defaults} with aliases like \alias{mv}{mv~-i}.
Here, the \cmd{mv} command is redefined to always run interactively, prompting the user at critical points, i.e. before overwriting existing files.
The default operating mode of \cmd{mv}, and most other commands, is to assume that the user is aware of and okay with the possible consequences of running it---and that they have not made any mistakes in its invocation.
This is of course a much more useful assumption in a scripting context.

The bias of most command-line tools towards scripting is also evident in their output, which is usually minimal and not tailored for human ease-of-use.
We can see this in aliases like \alias{mount}{mount~|~column -t}, which aligns the output of the \cmd{mount} command for easier reading, or \alias{df}{df~-h} or \alias{ll}{ls~-lh}, which change the default output of these commands so that file sizes are not shown simply in bytes but rather in much more practical common units like megabytes.
The high prevalence of aliases for \emph{colorizing output} (e.g. \alias{grep}{grep --color=auto}) is also notable, as color only makes sense in an interactive context.
In terminals, colorful text is achieved by inserting ANSI escape codes into the text stream.
This is a hindrance for scripts, but tools could easily detect whether they are run in an interactive terminal or as part of a script and adjust their output accordingly.

Note that the tension between interactivity and scripting is not the same as the divide between ``casual'' and ``power'' users.
Experts are experiencing the same frustrations as amateurs when using the shell interactively.
Recently, there has been a growing movement that sees today's command line as a \emph{human-first} text-based UI, rather than a \emph{machine-first} scripting platform \citep{clig}.
This new generation of command-line users and tool authors embrace the Unix philosophy with its core tenet of simple tools that can be composed well together \citep{raymond:03}, but they want to modernize those tools to fit current environments, with a more humanistic approach to their interaction design.\footnote{It should be noted that \citet{ritchie:74}, in a paper that pre-dates the invention of the Bourne shell, explicitly highlight the interactivity of the Unix system over the batch-processing nature of its predecessors. Today's notion of interactivity is of course more advanced, and it is now the classic Unix systems with their shell scripts that evoke an atmosphere of ``batch processing.''}
Emphasizing the conversational nature of the command line, they highlight the need for features such as error correction (cf. Section~\ref{sec:repair-rules}) or command suggestions (cf. Section~\ref{sec:discovering-workflows}), and confirming potentially destructive actions before they are executed.
They see human-readable output as paramount and suggest tools should be more aware of their environment (cf. Section~\ref{sec:contextual-defaults}).

\section{Threats to Validity}

We review potential limitations of our study as threats to validity.
First, our sample might not be representative.
Our dataset only includes aliases by people who publicly shared their dotfiles, we only collected from GitHub, and our sample does not include forks.
Nevertheless, our dataset is very exhaustive, as we were able to sample \per{94.09} of the estimated population of Shell files containing aliases on GitHub.
And while mining GitHub can be fraught with perils \citep{kalliamvakou:14}, we specifically sought out personal repositories, side-stepping many of the typical issues with mining GitHub for software projects.

Second, our parser might not be sophisticated enough to recognize complex real-world aliases or cope with minute platform differences.
To mitigate this threat, we ran multiple sanity checks and tested the parser on some hairy examples from the dataset.
We did not detect any significant mis-parses and think that we have covered the majority of relevant cases.
The raw unparsed database is available in our replication package.

Third, aliases might not reflect intent as much as we assume.
En-masse copy-pasting of aliases by users, without them knowing exactly what they are copying, is certainly a realistic scenario.
System distributions and configuration frameworks like \emph{ohmyzsh} ship with numerous aliases by default or as part of easily enabled plugins.
Users might not even be aware of the aliases they have on their system.
We mitigate this concern by removing all duplicate files from our dataset that would indicate sheer copy/pasting.
%We also particularly exclude alias definitions that come bundled with operating systems and particular shells (e.g., zsh).
%This is mitigated by the fact that all of the aliases we collected were publicly shared by users.
%It stands to reason that even if a user is not aware of all the details of their system configuration, they confirm their attachment to this configuration and its aliases by publicly sharing them---even if only for the purposes of synchronizing them across the users' own machines.

Fourth, we might not actually be able to see the true user intent, if it exists, as quantitative measures might hide a long tail of minor variations and individual user preference.
Conclusions about common aliases or selected subsets might not be generalizable.
To mitigate these summarizing effects, we established customization practices as a vehicle to take a deeper dive into the details of certain alias usage.
Since we sampled almost the whole available population, we are confident in the strength of our data and the conclusions we can draw from particular instances.
Our replication package includes our whole toolchain and all alias data in a relational format ready for further analysis.

\section{Related Work}

Related research in the broader context of our work has been conducted on understanding common practices in the software engineering community based on public online data, on software configuration in general, on the use of command-line interfaces and how to improve them, and on the shell as a programming language for both scripting and interactive use.

Empirical studies similar to ours, looking at community knowledge in software engineering to understand practices and distill insights, have been conducted in related domains:
\citet{zhong:15} study real-world bug fixes in Java projects to help guide automatic program repair;
\citet{yang:17} mine Stack Overflow posts and GitHub repositories to find out how programmers use and adapt copy-pasted code snippets in open-source projects, while \citet{baltes:19} investigate to what extent such snippets are copied without proper attribution;
\citet{prana:19} conduct a qualitative study to categorize the content of GitHub README files and build an automated classifier to label README sections, easing information discovery;
\citet{barnaby:20} present a tool that mines code bases for idiomatic usage examples of API methods.

In the context of software configuration, \citet{sayagh2020survey} surveyed experts and the literature to identify a number of challenges and recommendations related to configuration practices.
Our work reflects some of their findings, insofar as shell aliases are a form of personal configuration that can interact with---and counteract---other system configurations.
For example, selecting good out-of-the-box default values is seen as an important issue by experts, and aliases are indeed often used to \emph{override defaults}.
Related to our implications on contextual defaults (Section~\ref{sec:contextual-defaults}), \citet{zhen2011massconf} present MassConf, a system that proposes optimal software configurations based on a user's environment and existing configurations.
Adjacent work in configuration mining includes the ConfigMiner tool by \citet{sayagh2020configminer}, which identifies appropriate configuration options based on related StackOverflow questions.

The earliest study we found on the use of command-line interfaces was by \citet{greenberg:88a}, who collected four months of continuous real-life use of the Unix \verb|csh| shell from 168 users. 
The data was used in a follow up study to analyze the use of interactive systems by examining the frequency of command invocations for different groups of users~\citep{greenberg:88b}.
In later work, \citet{davison:98} use probabilistic action modeling to predict user action sequences based on the same dataset.
\citet{korvemaker:00} similarly predict future action sequences in command lines, but condition on actions of the particular user group with the goal of enabling adaptive user interfaces.
Other work in the context of adaptive user interfaces by \citet{jacobs:01} uses association rule learning on the shell logs to produce scripts to automate common task sequences.
\citet{khosmood:14} use the same corpus and two additional, more recent, corpora to learn a model that can identify user profiles based on their command-line behavior.
Bespoke~\citep{bespoke:19} is a system that synthesizes specialized graphical user interfaces (GUIs) based on command usage.
Our work can be viewed as an input to this system that passes common shell workflows in aliases to be generated as GUIs.

There has been other work on enhancing user experience in command-line interfaces.
NoFAQ~\citep{dantoni:17} provides repair suggestions for failed shell invocations based on a model learned from a curated set of fix patterns.
NL2Bash~\citep{nl2bash} implements a system that translates natural language phrases in English to shell commands.
Recent work by \citet{greenberg:17} has been looking into understanding the POSIX shell as a programming language.
More specifically, understanding word expansion in the shell to support interactivity~\citep{greenberg:18a} and concurrency~\citep{greenberg:18b}.

\section{Conclusion}

We report on a large-scale exploratory study on how command-line users customize user experience by defining shell aliases.
Through inductive coding, nine customization practices emerged from our dataset of collective customization knowledge mined from GitHub, providing insight on the characteristics of command-line use.
Based on our results, we discuss and formulate a set of implications for command-line tool developers, researchers, and the shell as an interactive environment for experts.
We enable further analysis and a basis for learning applications based on our extensive curated dataset.

Aliases often redefine commands with other default arguments, which is a potential indicator for usability problems in these tools.
However, we have to also be aware that defaults can be highly contextual depending on user profiles (e.g., expertise level) and environment (e.g., scripting vs. interactive use).
We also see our dataset and results as a rich source for learning norms with respect to repair rules, data flows, and descriptive names for complex command structures.
We provide a comprehensive replication package and see potential for future work based on our dataset and analyses.

\bibliographystyle{ACM-Reference-Format}
\bibliography{references}

\end{document}